\documentclass[twocolumn]{aastex63}

\usepackage{tikz}
\usepackage[english]{babel}
\usepackage[utf8]{inputenc}
\usepackage[colorinlistoftodos, color=green!40, prependcaption]{todonotes}
\usepackage{subfigure}

\newcommand{\YTeV}{Y_{\rm TeV}}
\newcommand{\LTeV}{L_{\rm TeV}}
\newcommand{\LMaxTeV}{L_{\rm TeV,\, Max}}
\newcommand{\LMinTeV}{L_{\rm TeV,\, Min}}

\newcommand{\PhiTeV}{\Phi_{\rm TeV}}
\newcommand{\PhiTeVTh}{\Phi_{\rm TeV}^{\rm th}}

\usepackage{lineno}

\begin{document}

\title{Unresolved sources naturally contribute to PeV $\gamma$-ray  diffuse emission observed by Tibet AS$\gamma$}

\author{V. Vecchiotti}
\affiliation{Gran Sasso Science Institute, 67100 L'Aquila, Italy}
\affiliation{INFN, Laboratori Nazionali del Gran Sasso, 67100 Assergi (AQ),  Italy}

\author{F. Zuccarini}
\affiliation{University of L'Aquila, Physics and Chemistry Department, 67100 L'Aquila, Italy}
\affiliation{INFN, Laboratori Nazionali del Gran Sasso, 67100 Assergi (AQ),  Italy}

\author{F.L. Villante}
\affiliation{University of L'Aquila, Physics and Chemistry Department, 67100 L'Aquila, Italy}
\affiliation{INFN, Laboratori Nazionali del Gran Sasso, 67100 Assergi (AQ),  Italy}

\correspondingauthor{Giulia Pagliaroli}
\email{giulia.pagliaroli@gssi.it}
\author{G. Pagliaroli}
\affiliation{Gran Sasso Science Institute, 67100 L'Aquila, Italy}
\affiliation{INFN, Laboratori Nazionali del Gran Sasso, 67100 Assergi (AQ),  Italy}

\begin{abstract}
The Tibet AS$\gamma$ experiment provided the first measurement of the total diffuse gamma-ray emission from the Galactic disk in the sub-PeV energy range. Based on the analysis of TeV sources included in the HGPS catalog, we predict the expected contribution of unresolved pulsar-powered sources in the two angular windows of the Galactic plane observed by Tibet AS$\gamma$. 
We show that the sum of this additional diffuse component due to unresolved sources and the truly diffuse emission, due to cosmic ray interaction with the interstellar medium, well saturates the Tibet data, without the need to introduce a progressive hardening of the cosmic-ray spectrum toward the Galactic center.
\end{abstract}



\section{Introduction} \label{sec:outline}

The Tibet AS$\gamma$ collaboration has recently obtained a measurement of the Galactic diffuse $\gamma$-ray emission, providing the first
evidence that this component extends to 
sub-PeV energies \citep{TibetASgamma:2021tpz}.
The origin of this high-energy diffuse emission is extensively debated in the very recent literature, see e.g. \cite{Liu:2021lxk,Koldobskiy:2021cxt,Fang:2021ylv,Neronov:2021ezg}.

%
An essential step for the interpretation of Tibet AS$\gamma$ data is the evaluation of the cumulative flux produced by sources that are too faint to be individually resolved. 
These give rise to a large-scale diffuse emission, superimposed to that produced by cosmic ray (CR) interactions with the interstellar medium, that can affect the space and energy distribution of the observed signal.
Population studies of TeV Galactic sources have shown that unresolved contribution is not negligible in the TeV domain, see e.g. \cite{Cataldo:2020qla,Steppa:2020qwe}. 
Recently Tibet AS$\gamma$ \citep{TibetCRAB} and HAWC \citep{HAWC:GalacticHE} have shown that several Galactic sources produce $\gamma-$rays above $\sim 50$~TeV.
Moreover, LHAASO-KM2A reports the detection of more than $530$ photons at energies above $100\;{\rm TeV}$ and up to $1.4\;{\rm PeV}$ from 12 ultra-high-energy $\gamma$-ray sources \citep{2021Natur.594...33C}.
It is thus natural to expect that unresolved source contribution may be relevant also in the sub-PeV energy range.

In this paper, we took advantage of the observations provided by H.E.S.S. Galactic Plane Survey (HGPS) in the $1-100$~TeV energy domain to discuss the implications of unresolved pulsar-powered sources for the interpretation of Tibet AS$\gamma$ results.
Our approach is based on standard assumptions for the spatial and intrinsic luminosity source distributions (also employed e.g. \cite{Steppa:2020qwe,Strong:2006hf,Pothast:2018bvh}).
These prescriptions are physically well-motivated for a population of sources powered by pulsar activity, like Pulsar Wind Nebulae (PWNe) or TeV halos which could represent the dominant component of the TeV sky, as recently suggested by \cite{Sudoh:2019lav,Abdalla:2017vci}. 
We show that the inclusion of unresolved contribution from these sources, adding up to standard predictions for CR diffuse flux \citep{Lipari:2018gzn}, naturally explains the Tibet AS$\gamma$ data, without the need to introduce a progressive hardening  of the CR spectrum toward the Galactic center.

%
%

\section{Source population}

It is naturally expected that sources provide a relevant contribution to high energy $\gamma$-ray  Galactic emission, see e.g. \cite{Ahlers:2013xia} for an estimate of the total emission expected from different classes of Galactic sources. %
However, the key point is to quantify the fraction of sources that are not resolved by experiments, since these generate a cumulative diffuse emission that adds up to the one produced by CR interactions. 
The goal of this paper is to predict the unresolved pulsar-powered sources contribution to galactic $\gamma$-ray emission in TeV and sub-PeV energy domain. 
In order to estimate this quantity, we consider the information provided by HGPS in the energy range $1\, {\rm TeV} \le E_\gamma \le 100\,{\rm TeV}$.

We assume that average source emission spectrum can be described by a power-law with exponential cutoff, i.e. $\varphi(E_\gamma) \propto E_{\gamma}^{-\beta} \, \exp\left(-E_{\gamma}/E_{\rm cut}\right)$.
The index $\beta$ is given by the average spectral index of sources included in the HGPS catalog, i.e. $\beta = 2.3$.
The cutoff energy is chosen as $E_{\rm cut} = 500\,{\rm TeV}$.
This value is not yet constrained and it corresponds to assuming that TeV source spectrum observed by H.E.S.S. can be extrapolated with a moderate suppression in the sub-PeV region. 
In this respect, it is interesting to note that the Tibet AS$\gamma$
experiment has recently shown that Crab Nebula emits $\gamma-$rays in
the sub-PeV region with $5.6\sigma$ statistical significance \citep{TibetCRAB}. 
Moreover, HAWC experiment has reported evidence of several Galactic
sources emitting above $\sim 50\,{\rm TeV}$ \citep{HAWC:GalacticHE}
whose $\gamma$-ray flux can be described as leptonic emission from
electrons/positrons injected with power-law spectrum and exponential cutoff around 1 PeV \citep{Sudoh:2021avj}.
We remark that the adopted value of $E_{\rm cut}$ moderately affects
flux predictions for the low energy data points given by Tibet
AS$\gamma$ that are relatively close to the range probed by H.E.S.S..
At larger energies ($E_{\gamma} \ge 1\; {\rm PeV}$), the source emission spectrum is limited by photons absorption in the interstellar radiation field (mainly due to Cosmic Microwave Background radiation) that suppresses the flux produced by distant sources.
We take this into account, as it is described in \cite{Vernetto:2016alq}.
The sources space and luminosity distribution is described by:
\begin{equation}
\frac{dN}{d^3 r\,d\LTeV} = \rho\left({\bf r} \right) \YTeV \left(\LTeV\right)  
\label{SpaceLumDist}
\end{equation}
where $r$ indicates the distance from the Galactic Center. 
The function $\rho({\bf r})$ is assumed to be proportional to the pulsar distribution parameterized by \cite{Lorimer:2006qs} and to scale as $\exp \left(-\left|z  \right|/H\right)$ with $H=0.2\ {\rm kpc}$, along the direction $z$ perpendicular to the Galactic plane.
It is conventionally normalized to one when integrated in the entire Galaxy.

The function $\YTeV(\LTeV)$ gives the source intrinsic luminosity distribution in the TeV energy domain\footnote{We indicate with $\LTeV$ and $\PhiTeV$ the integrated source luminosity and flux and in the energy range $1-100\;{\rm TeV}$ probed by H.E.S.S.}. 
It is parameterized as a power-law:
\begin{equation}
\YTeV(\LTeV)=\frac{{R \, \tau \, (\alpha-1)}}{\LMaxTeV}\left(\frac{\LTeV}{\LMaxTeV}\right)^{-\alpha}
\label{LumDist1} 
\end{equation}
in the luminosity range $\LMinTeV\le\LTeV \le
\LMaxTeV$.
This distribution is naturally obtained for a population of {\em  fading} sources, such as PWNe or TeV Halos, created at a constant rate $R$ and having
intrinsic luminosity that decreases over a time scale $\tau$ according to:
\begin{equation}
\LTeV(t)= \LMaxTeV\left(1+\frac{t}{\tau}\right)^{-\gamma}
\label{lum}
\end{equation}
where $t$ indicates the time passed since source formation.
In this assumption, the exponent $\alpha$ of the luminosity distribution is given by $\alpha = 1/\gamma + 1$.

The birth rate of PWNe or TeV Halos is similar to that of SN explosions in our
Galaxy, i.e. $R\simeq R_{\rm SN} = 0.019\,{\rm yr}^{-1}$ \citep{Diehl:2006cf}. 
Since $\gamma-$ray emission is powered by pulsar activity, the TeV-luminosity $\LTeV$ is a fraction of the spin-down power $\dot{E}$, i.e. we assume  $\LTeV = \lambda\, \dot{E}$ where $\lambda \le 1$. 
If pulsar energy loss is dominated by magnetic dipole radiation (braking index $n=3$) and the efficiency of TeV emission does not depend on time ($\lambda\sim {\rm const}$), the exponent in Eq.~(\ref{lum}) is $\gamma=2$, that corresponds to a source luminosity function $\YTeV(\LTeV)\propto \LTeV^{-1.5}$.
The possibility of $\lambda$ being correlated to the spin-down power, i.e. $\lambda = 
\lambda_0 ({\dot E}/{\dot E_0})^{\delta}$, was suggested by \cite{Abdalla:2017vci} that found $\LTeV = \lambda\,{\dot E} \propto \dot{E}^{1+\delta}$ with $1 + \delta = 0.59 \pm 0.21$ by studying a sample of PWNe in the HPGS catalog\footnote{ Our predictions are not sensitive to the absolute value for $\lambda_0$ being it absorbed inside the parameter $\LMaxTeV=\lambda_{0} \dot{E}_{0}$ that we obtain as a best-fit from the HGPS data.}. In this case, one obtains $\gamma \simeq 1.2$ in Eq.~(\ref{lum}) that corresponds to a source luminosity function $\YTeV(\LTeV)\propto \LTeV^{-1.8}$.

\section{The unresolved source contribution}

In a recent paper \citep{Cataldo:2020qla}, we have shown that
H.E.S.S. observations can be used to efficiently constrain the TeV sources population. 
In particular, we considered HGPS sources producing a photon flux above 1 TeV larger than $10\%$ of the CRAB flux. Above this threshold, the HGPS catalog can be considered complete \citep{H.E.S.S.:2018zkf} and it includes $32$ sources. We removed from the analysis sources firmly associated with SNRs (i.e. Vela Junior, RCW 86, RX J1713.7-3946) and we treated the residual $29$ sources as pulsar-powered. This assumption is justified being this subset composed by $8$ firmly identified PWNe, $2$ composite objects, showing evidence of both shell and nebular emission, and $19$ unidentified sources. In the unidentified sub-sample, $12$ sources have been considered as candidate PWNe in further studies on the basis of new data and/or phenomenological considerations (\cite{Sudoh:2021avj}, \cite{2008ICRC....3.1341W}, \cite{Abdalla:2017vci}, \cite{Giacinti:2019nbu}). The average spectral index of the considered sample of 29 HGPS sources is $2.34$ supporting our assumption $\beta = 2.3$.

Namely, by fitting the flux, latitude and longitude distribution of
bright sources in the HGPS catalog, we have obtained $\LMaxTeV =  4.9^{+3.0}_{-2.1} \times 10^{35} {\rm erg\;s^{-1}}$
and $\tau =  1.8^{+1.5}_{-0.6}\times 10^3\,{\rm yr}$ for $\alpha=1.5$ ($\LMaxTeV=6.8^{+6.7}_{-3.0}\times 10^{35} {\rm erg\;s^{-1}}$ and $\tau = 0.5^{+0.4}_{-0.2} \times 10^3\,{\rm yr}$
for $\alpha=1.8$). These results are also valid for extended TeV sources, provided that they have dimensions that do not exceed $\sim 40$ pc.
Moreover, our results are consistent with those obtained by a completely independent analysis of HGPS catalog performed by \cite{Steppa:2020qwe}. 
In this paper, we take advantage of the results of \cite{Cataldo:2020qla} to estimate the unresolved flux produced by the considered population in the TeV and sub-PeV energy domain. 
For definiteness, we take as a reference the case $\alpha =1.5$. 
Slightly larger fluxes are obtained for $\alpha=1.8$.

In order to evaluate the cumulative contribution to diffuse $\gamma-$ray signal of sources which are
too faint to be individually detected, we introduce a flux
detection threshold $\PhiTeVTh$ whose value is estimated by considering
the performances of H.E.S.S. detector. 
H.E.S.S. is able to resolve point-like sources if they produce an integrated flux $\PhiTeV$ in the $[1-100]\;{\rm TeV}$ energy domain that is larger than $0.01 \Phi_{\rm CRAB}$, where $\Phi_{\rm CRAB} = 2.26\times 10^{-11}\;{\rm cm^{-2}}\;{\rm s^{-1}}$ is the flux produced by CRAB in the same energy range.   
Extended sources can however escape detection even if their flux
exceeds this value, as it is e.g. understood by looking the sensitivity curve
presented in Fig.~13 of \cite{H.E.S.S.:2018zkf}.  
The HGPS catalog can be considered complete for objects producing fluxes larger than $0.1 \Phi_{\rm CRAB}$ (with the exception of sources having an angular extension larger than $\sim 1^\circ$ which cannot be observed by H.E.S.S).  
Taking this into account, we calculate the cumulative emission of unresolved sources as:
\vspace{-0.14cm}
\begin{eqnarray}
\nonumber
\varphi^{\rm NR}(E_\gamma) &=& 
\varphi(E_\gamma) \; \eta(E_\gamma,\PhiTeVTh) \,\times \\
&\times&
\int_{0}^{\PhiTeVTh} d\PhiTeV\; \PhiTeV\,
\frac{dN}{d\PhiTeV}
\end{eqnarray}
where $\varphi(E_\gamma)$ is the average source emission spectrum\footnote{
The constant $K$ is determined by the condition that
$\varphi(E_\gamma)$ is normalized to one when integrated in the
$[1-100]\;{\rm TeV}$ energy domain, i.e. $K\equiv \int_{\rm 1\, TeV}
^{\rm 100\, TeV}
dE_\gamma \; \varphi(E_\gamma)$.}: 
\begin{equation}
\varphi(E_\gamma) = \frac{1}{K} \left(\frac{E_\gamma}{1\, {\rm  TeV}} \right)^{-\beta}
\end{equation}
the quantity $dN/d\PhiTeV$ is the source flux distribution in a given region of the sky (see  Eq.~A.6 of \cite{Cataldo:2020qla}).  The function $\eta(E_\gamma,\PhiTeVTh)$ is the average survival probability of photons with energy $E_\gamma$ emitted by unresolved sources. It includes the gamma absorption due to gamma-gamma interactions on CMB photons following the numerical approach described in \cite{Lipari:2018gzn}. We vary the flux detection threshold in the range $\PhiTeVTh = [0.01 - 0.1] \Phi_{\rm CRAB}$.

Note that the diffuse Galactic $\gamma$-ray flux at
sub-PeV energies measured by Tibet AS$\gamma$ is obtained by subtracting/masking the contribution of sources which are included
in the TeVCAT catalog \citep{2008ICRC....3.1341W}. 
This implies that sources should be faint not only at sub-PeV
energies (but also at TeV energies) to escape detection.
As a consequence, the above approach which is based on the detection capabilities of
experiments operating in the TeV domain is also adequate to
investigate unresolved source contribution in the sub-PeV energy range.

%
\begin{figure}[h!]
\includegraphics[width=0.45\textwidth]{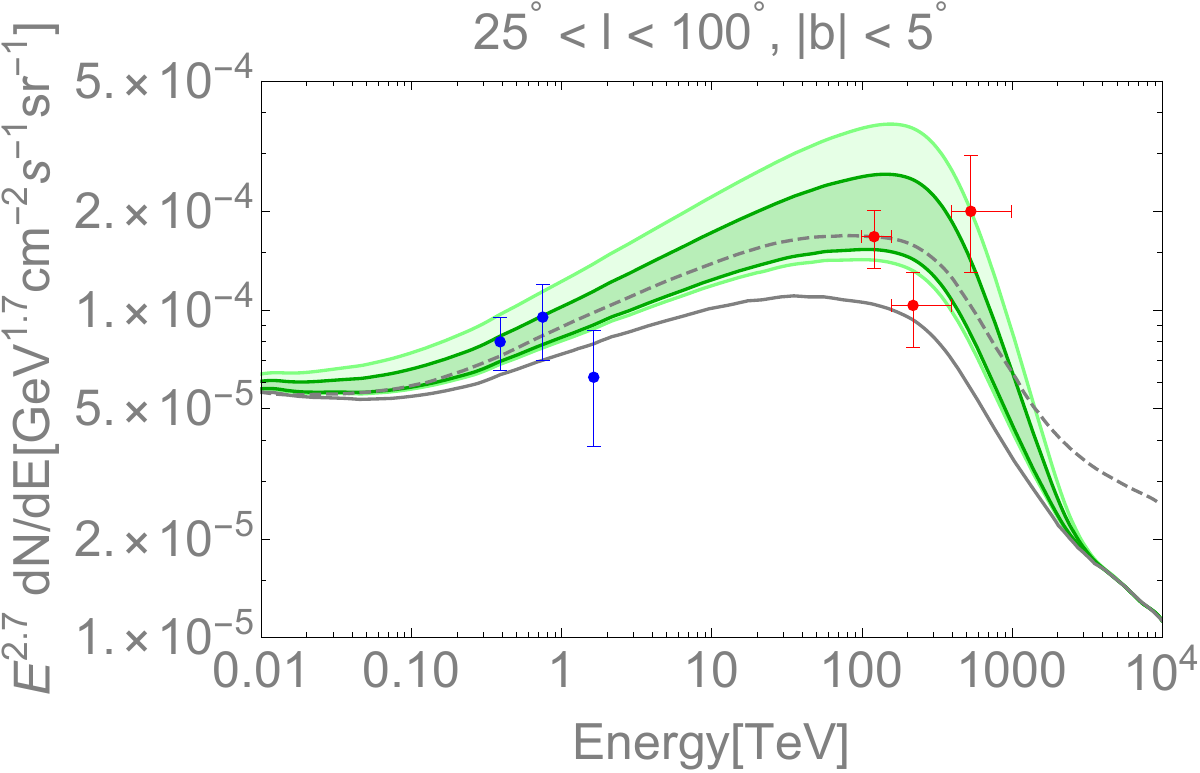}\\
\\
\includegraphics[width=0.45\textwidth]{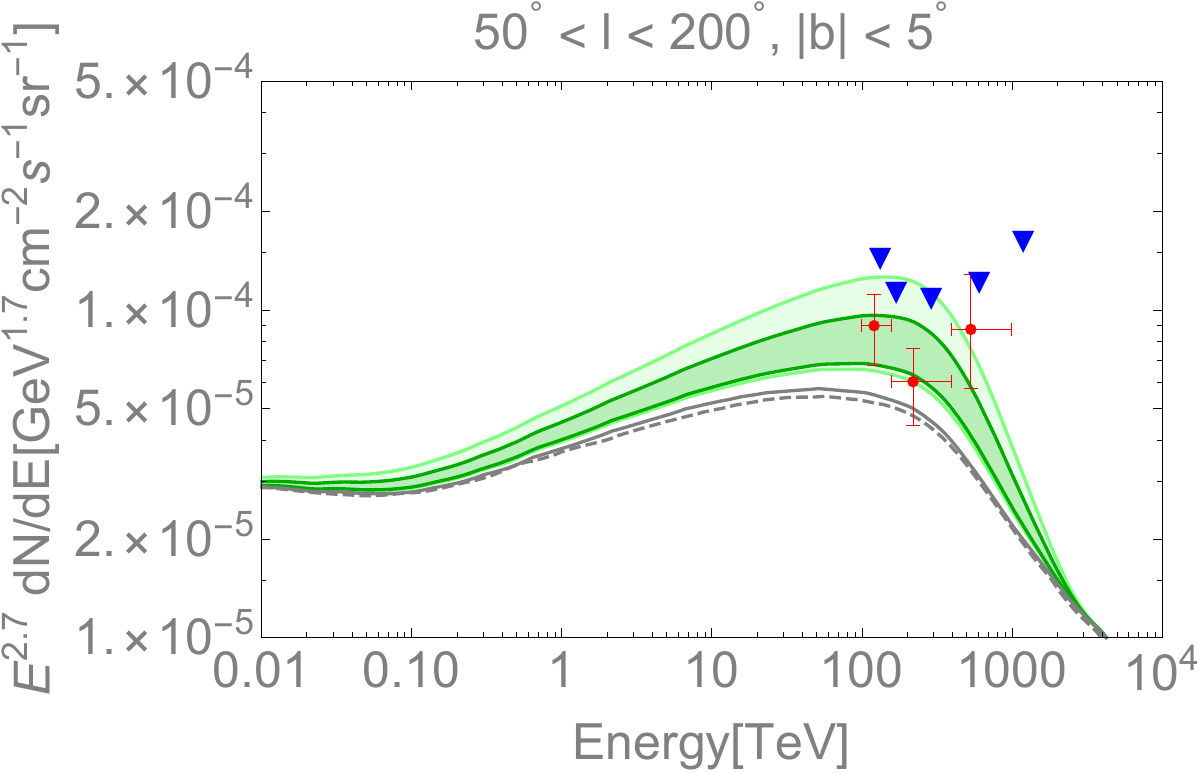}
\caption{\small\em Differential energy spectra of diffuse $\gamma-$rays from the Galactic plane in two different angular regions. Red data points are the measurements provided by Tibet \citep{TibetASgamma:2021tpz}. Blue data points in the upper panel are provided by Argo-YBJ \citep{ARGO-YBJ:2015cpa}, while blue triangles in the lower panel are upper limits by the CASA-MIA experiment \citep{CASA-MIA:1997tns}. Solid and dashed curves display the predicted energy spectra by the space-independent and space-dependent models by \cite{Lipari:2018gzn}, respectively. The green shaded band represents the total diffuse $\gamma-$ray emission obtained by adding the unresolved source contribution estimated in this paper to the $\gamma$-ray truly diffuse emission from space-independent model by \cite{Lipari:2018gzn}.   
}
\label{fig:tibet1}
\end{figure}

\section{Results}
\label{sec:Results}

The flux produced by faint sources which are not individually detected adds up the CR diffuse emission, shaping the radial and spectral behaviour of the total diffuse $\gamma$-ray flux observed by different experiments.
In Fig.~\ref{fig:tibet1}, we shows the theoretical predictions for
the {\em total} diffuse $\gamma-$ray flux (green band) as a function of energy in the regions of the Sky probed by Tibet
AS$\gamma$ experiment \citep{TibetASgamma:2021tpz}.
The upper panel (a) refers to the region $25^{\circ}<l<100^{\circ}$,
while the lower panel (b) shows the region $50^{\circ}<l<200^{\circ}$;
both of them correspond to the latitude range $|b|<5^\circ$.
The unresolved sources contribution is obtained as described in the
previous section and the thickness of the darker green band corresponds to the uncertainty in flux detection threshold $\PhiTeVTh$. 
Namely, the upper and lower green lines are obtained by
assuming $\PhiTeVTh = 0.1\; \Phi_{\rm CRAB}$ and $\PhiTeVTh = 0.01\;\Phi_{\rm CRAB}$, respectively.
The light green band also includes the uncertainty on this prediction due to the correlated variations of the source population parameters $\tau$ and $\LMaxTeV$ within their $1\sigma$ uncertainty.
The truly diffuse emission, produced by CR interactions with the
interstellar gas, is shown by grey solid lines in 
Fig.\ref{fig:tibet1} and corresponds to the ``space-independent'' model
of \cite{Lipari:2018gzn}.
Red data points show the diffuse flux measured by Tibet AS$\gamma$. 
These are obtained after subtracting events within 0.5$^\circ$
from known TeV sources included in the TeVCAT catalog \citep{2008ICRC....3.1341W}.
The error bars show 1$\sigma$ statistical errors. 
Finally, we also display the CR diffuse
flux corresponding to the ``space-dependent'' model of  \cite{Lipari:2018gzn}
(gray dashed lines) to permit comparison with our predictions.
This is obtained by assuming that the CR spectrum in the inner Galaxy
is harder than at the Sun position, as it seems to be suggested by Fermi-LAT
data \citep{Acero:2016qlg,Yang:2016jda, Pothast:2018bvh} (see
\cite{Vecchiotti:2021vxp}, \cite{Peron:2021lwb} for alternative explanations)\footnote{The implications of CR spectral hardening in the inner Galaxy have been also discussed in \cite{Pagliaroli:2016lgg,Pagliaroli:2017fse, Cataldo:2019qnz} by using a phenomenological approach proposed in \cite{Pagliaroli:2016lgg} similar to that adopted by \cite{Lipari:2018gzn}.}.

Several important conclusions can be obtained from Fig.~\ref{fig:tibet1}.
First, we see that unresolved source contribution is not negligible for
$E_\gamma\ge 1$ TeV and becomes progressively more relevant as 
energy increases as it was also pointed out in \cite{Linden:2017blp}.
This is a natural consequence of the fact that sources
are expected to have, on average, harder spectrum (a part from
cutoff effects) than CR diffuse emission. 
At the energy $E_\gamma \simeq 150 \;{\rm TeV}$, corresponding to the
first data point of Tibet AS$\gamma$, the cumulative flux produced by faint
sources is estimated to be $49\% - 154\%$ ($25\% - 79\%$) of the
truly diffuse signal in the region $25^{\circ}<l<100^{\circ}$ ($50^{\circ}<l<200^{\circ}$). 
%

It is useful to investigate the stability of our results concerning the possibility that some of the 29 sources considered in our analysis turn out to be not pulsar-powered. For this purpose, 
we repeat the fit of the flux, latitude and longitude distribution of HGPS sources taking into account only the 22 sources with a clear or potential association to PWNe. The new best-fit values with this reduced sample are $L_{\rm TeV,\, Max} =6.2\times 10^{35}{\rm \;erg\;s^{-1}}$ and $\tau = 1.1 \times 10^3\,{\rm yr}$. The unresolved flux due to the PWNe population obviously decreases, however, the prediction still falls inside the $1\sigma$ statistical uncertainty band reported in light green in Fig.~\ref{fig:tibet1}.

\begin{figure}[h!]
\includegraphics[width=0.45\textwidth]{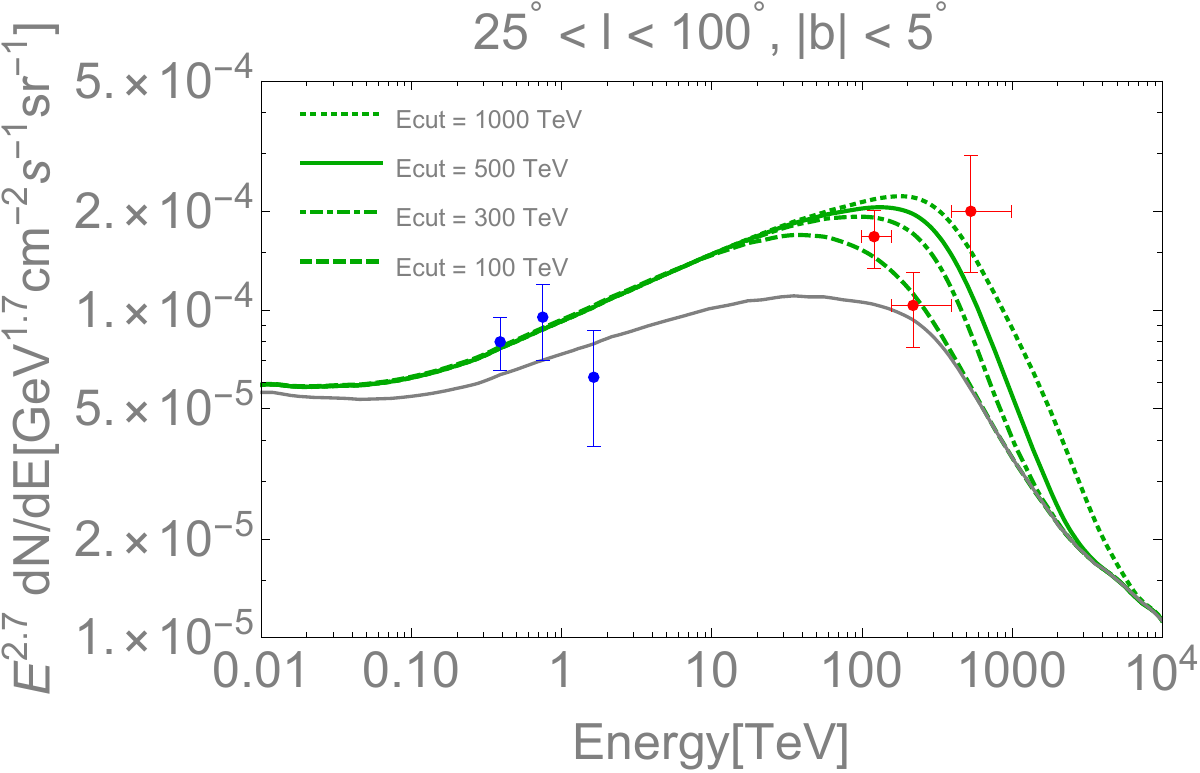}\\
\\
\includegraphics[width=0.45\textwidth]{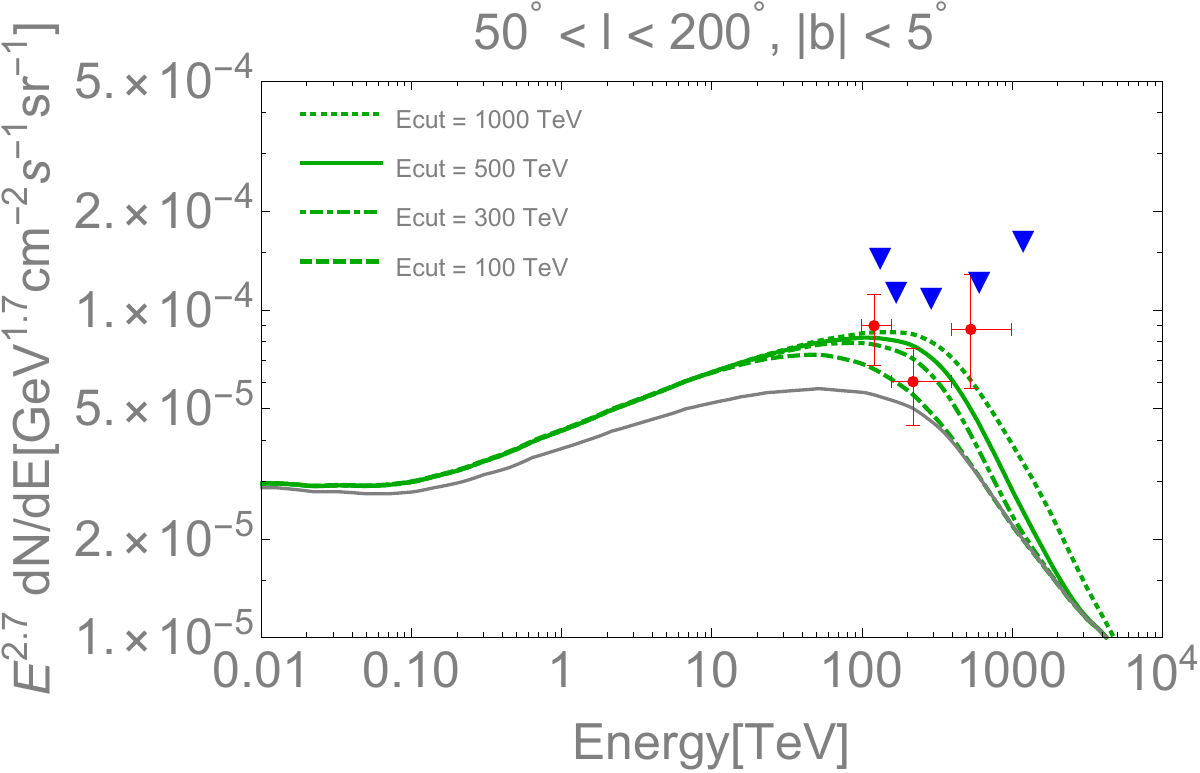}
\caption{\small\em 
We highlight the role of the assumed energy cut for the sources spectra.
The green lines show the differential energy spectra of diffuse $\gamma-$rays from the Galactic plane in two different angular regions for an intermediate sensitivity threshold of $\PhiTeVTh\sim 0.05\,\Phi_{\rm CRAB}$ and assuming different values for the spectral energy cut-off from $1$ PeV to $100$ TeV.   
\label{fig:tibet2}}
\end{figure}

We remark that our prediction is only marginally dependent on the high-energy extrapolation of the source spectra (e.g. the adopted value of the cutoff energy) since the lowest energy data point of Tibet AS$\gamma$ experiment nearly overlaps with the energy range probed by H.E.S.S..
This is evident in Fig.\ref{fig:tibet2} where green lines represent the total diffuse $\gamma$-ray emission for intermediate sensitivity threshold $\PhiTeVTh\sim 0.05\,\Phi_{\rm CRAB}$, obtained by assuming different values for the spectral energy cut-off from $1$ PeV to $100$ TeV.
In particular, if we assume a lower energy cutoff $E_{\rm cut}=300$ TeV ($E_{\rm cut}=100$ TeV), the unresolved source flux at $150$~TeV decreases by $\sim 17\%$ ($\sim 68\%$) with respect to the reference case $E_{\rm cut}=500$ TeV. 
%
%
We can also consider the effects of source spectral index variations (not shown in Fig.~\ref{fig:tibet2} to avoid overcrowding it). 
If we take $\beta=2.4$ ($\beta=2.5$ TeV), the unresolved source flux at $150$~TeV decreases by $\sim 32\%$ ($\sim 54\%$) with respect to the reference assumption $\beta=2.3$.
In all these cases, our predictions for the total diffuse $\gamma$-ray emission are still consistent with the first two Tibet data points.

%

\begin{figure}[h!]
\begin{center}
\includegraphics[width=0.4\textwidth]{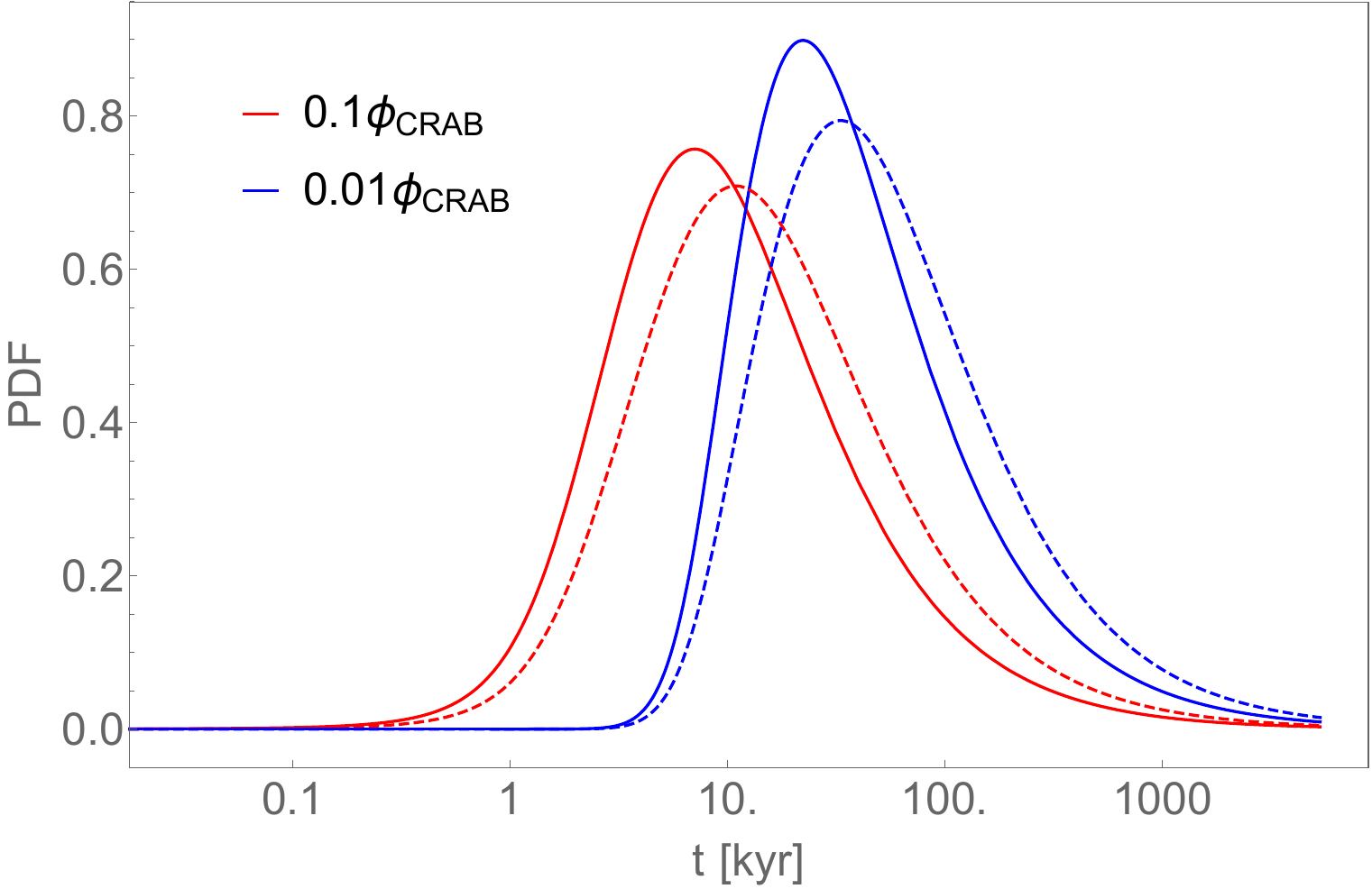}
\caption{\small\em The distribution of sources contributing to unresolved signal as a function of  $\log_{10}(t/ 1\,{\rm kyr})$. Thick lines refer to the sky region $|b|<5^{\circ}$ and $25^{\circ}<l<100^{\circ}$ while dashed ones refer to the region between $50^{\circ}<l<200^{\circ}$. The blue (red) lines are obtained for $\PhiTeVTh=0.01 \Phi_{\rm CRAB}$ ($\PhiTeVTh=0.1 \Phi_{\rm CRAB}$).
\label{fig:tibet3}}
\end{center}
\end{figure}

It is interesting to investigate the typical age of sources that give relevant contribution to unresolved signal.
In Fig.\ref{fig:tibet3} we show the relative contribution to unresolved emission as a function of $\log_{10}\left(t/{\rm 1 kyr}\right)$ where $t$ is the age of the source, i.e. the time passed since pulsar formation.
Thick lines refer to the sky region $|b|<5^{\circ}$ and $25^{\circ}<l<100^{\circ}$ while dashed ones refer to the most lateral region between $50^{\circ}<l<200^{\circ}$. With blue lines we show the case of detection threshold $\PhiTeVTh=0.01 \Phi_{\rm CRAB}$, corresponding to the lower bound of the green band for the total diffuse flux reported in Fig.\ref{fig:tibet3}. In this case the dominant contribution is provided by PWNe with an age ranging between $t\sim (22-33)$ kyr, depending on the sky region considered. Red lines correspond to  
the upper bound of the green band for the total diffuse flux reported in Fig.\ref{fig:tibet1} with $\PhiTeVTh=0.1 \Phi_{\rm CRAB}$. In this case a younger population provides a dominant contribution peaking between $t\sim (7-11)$ kyr, depending on the sky region considered. The unresolved flux contribution due to sources older than $\sim 100$ kyr (likely TeV Halos) is expected to be at most the $20\%$.  

It is important to remark that our calculations naturally reproduce the Tibet AS$\gamma$ results both in the low and high longitude observation
window, corroborating the conclusion that unresolved PWNe provide a
relevant contribution in this energy range. 
 A test of our calculations could be provided by additional observations in the TeV and PeV energy domain.
Future experiments with improved sensitivity will be e.g. able to resolve more sources in the regions of interest. 
As an example, CTA, with a sensitivity according to \cite{Sudoh:2019lav}, should be able to resolve about $280$ ($140$) pulsar-powered sources in the whole Galaxy, if the typical source size is $10$~pc ($40$~pc).

We also note that LHAASO-KM2A experiment recently presented preliminary determinations of the diffuse $\gamma-$ray signal in the sub-PeV energy domain \citep{Zhao:2021GJ}. 
%
%
%
The LHAASO-KM2A diffuse flux measurement is obtained by masking sources included in TeVCAT catalog
\citep{2008ICRC....3.1341W} with a relatively large mask
radius\footnote{The adopted mask radius is $R=2\sqrt{\sigma_{\rm ext}^2 + {\rm p.s.f.}^2}$,  where $\sigma_{\rm ext}$ represents the extension of the source while ${\rm p.s.f.}$ is LHAASO-KM2A point-spread-function, and it is typically comparable to or larger than $1^\circ$ \citep{Zhao:2021GJ}.}.
LHAASO-KM2A preliminary data for the longitude window $25^\circ\le l \le 100^\circ$ are lower than Tibet
AS$\gamma$ results and are close to predictions for the
``space-dependent'' CR-diffuse emission of \cite{Lipari:2018gzn}.
They cannot be however directly compared with theoretical
determinations of the diffuse signal because the masking
procedure cuts out a large part of the galactic plane, as it is seen
by Fig.~2 of \cite{Zhao:2021GJ}, where most of the CR-diffuse signal and unresolved emission is produced. 
They thus naturally represent a lower limit to the total diffuse emission.

Finally, by comparing our predictions in panel a) with the
truly-diffuse flux  given by the  ``space-dependent model'' of
\cite{Lipari:2018gzn} (grey dashed lines), we see that the unresolved
sources can mimic, at relatively low galactic longitudes, the effects produced by CR-spectral hardening in the
inner Galaxy (and vice-versa). 
This is in agreement with what we have found in \cite{Vecchiotti:2021vxp}. 
%
%
By looking at panel b), we note, however, that the previous statement is not
valid in the longitude range $50^\circ \le l \le 200^\circ$ where
unresolved sources,  differently from CR-spectral hardening, 
produce an enhancement of the diffuse emission 
and provide a good description of the Tibet AS$\gamma$ data. 

The different behaviour has a direct explanation. 
The diffuse emission at $l\ge 60^{\circ}$ is mainly produced,
for geometrical reasons, at galactocentric distances comparable to or larger than the
distance of the Sun from the Galactic center.
As a consequence, it is not sensitive to variations of the CR
distribution in the inner Galaxy. 
On the contrary, it can be increased by unresolved sources, which are distributed in the whole
Galaxy, and thus produce a non vanishing contribution even at large latitudes. 
%
%
We remark that that this point is important because it provides the
possibility to distinguish between the two effects in present and
future experiments. 
At the moment, the inclusion of unresolved PWNe contribution produces a better
description of the Tibet AS$\gamma$ data than CR spectral hardening.  

%
%

%

\section{Acknowledgements \label{sec:acknowledgement}}

We thank Silvia Vernetto and Paolo Lipari for useful discussion.
The work of GP and FLV is partially supported by the research grant number 2017W4HA7S ''NAT-NET:
Neutrino and Astroparticle Theory Network'' under the program PRIN 2017 funded by the Italian Ministero dell'Istruzione, dell'Universita' e della Ricerca (MIUR).


\bibliography{bibliography}
\bibliographystyle{aasjournal}


\end{document}